\documentclass[11pt]{article}

\usepackage[top=2cm, bottom=2cm, left=2cm, right=2cm]{geometry}

\usepackage[noadjust]{cite}

\usepackage[cmex10]{amsmath}
\usepackage{amsbsy}

\usepackage{stackengine}
\usepackage{makecell}
\usepackage{boldline}
\setcellgapes{3pt}

\usepackage[latin2]{inputenc}
\usepackage{amsmath}
\usepackage{amsbsy}
\usepackage{bm}
\usepackage{amsthm,dcolumn,graphicx,multicol,fancyhdr}
\usepackage{latexsym}

\usepackage{algorithm}
\usepackage{algpseudocode}
\usepackage{varwidth}

\usepackage{tikz}
\usetikzlibrary{shapes, shadows, arrows}
\usepackage{url}

\usepackage{graphicx}
\usepackage{float}
\usepackage{stfloats}
\usepackage{wrapfig}

\usepackage{epstopdf}

\usepackage{amsthm,dcolumn,graphicx,multicol,fancyhdr}
\usepackage{latexsym}
\usepackage{graphicx}
\usepackage{ifthen}
\usepackage{rotating,amsfonts,color,amssymb,amsmath,amscd,array,enumerate,afterpage}

\usepackage{hyperref}
\usepackage{multirow}

\usepackage{enumitem}
\setlist{parsep=3pt,listparindent=\parindent}

\usepackage[caption=false,font=footnotesize]{subfig}

\usepackage{array}
\newcolumntype{L}[1]{>{\raggedright\let\newline\\\arraybackslash\hspace{0pt}}m{#1}}
\newcolumntype{C}[1]{>{\centering\let\newline\\\arraybackslash\hspace{0pt}}m{#1}}
\newcolumntype{R}[1]{>{\raggedleft\let\newline\\\arraybackslash\hspace{0pt}}m{#1}}

\usepackage{arydshln}

\theoremstyle{definition}

\newcommand{\babc}{\renewcommand{\labelenumi}{(\alph{enumi})}\begin{enumerate}}
\newcommand{\eabc}{\end{enumerate}}
\newcommand{\biii}{\renewcommand{\labelenumi}{(\roman{enumi})}\begin{enumerate}}
\newcommand{\eiii}{\end{enumerate}}

\newcommand{\beqn}{\begin{eqnarray*}}
\newcommand{\beq}{\begin{eqnarray}}
\newcommand{\eeqn}{\end{eqnarray*}}
\newcommand{\eeq}{\end{eqnarray}}

\newcommand{\ckboldon}[1]{#1}

\newcommand{\ckbold}[1]{%
 \ifthenelse{\isundefined{\ckboldon}}{#1}{ \textbf{#1} }
}

\makeatletter
\def\hlinewd#1{%
  \noalign{\ifnum0=`}\fi\hrule \@height #1 \futurelet
   \reserved@a\@xhline}
\makeatother

\usepackage{color, colortbl}
\usepackage{gensymb}
\definecolor{Gray}{gray}{0.9}
\definecolor{LightCyan}{rgb}{0.88,1,1}

\begin{document}
\date{}
\title{A Comprehensive Review of Wind Energy in Malaysia: Past, Present and Future Research Trends}

\author{Fuad Noman\footnote{Institute of Sustainable Energy, Universiti Tenaga National, Jalan Ikram-Uniten, 43000 Kajang, Selangor, Malaysia (e-mail: fuad.noman@uniten.edu.my)},
 Gamal Alkawsi, Dallatu Abbas, Ammar Alkahtani, Seih Kiong Tiong,\\ Janaka Ekanyake
}

\markboth{To be Submitted for review}%
{Shell \MakeLowercase{\textit{et al.}}: Bare Demo of IEEEtran.cls for Journals}

\maketitle

\begin{abstract}
Wind energy has gained a huge interest in the recent years in various countries due to the high demand of energy and the shortage of traditional electricity sources. This is because it is cost effective and environmentally friendly source that could contribute significantly to the reduction of the ever-increasing carbon emissions. Wind energy is one of the fastest growing green technology worldwide with a total generation share of 564 GW as the end of 2018. In Malaysia, wind energy has been a topic of interest in both academia and green energy industry. In this paper, the current status of wind energy research in Malaysia is reviewed. Different contributing factors such as potentiality and assessments, wind speed and direction modelling, wind prediction and spatial mapping, and optimal sizing of wind farms are extensively discussed. This paper discusses the progress of all wind studies and presents conclusions and recommendations to improve wind research in Malaysia.\\
\end{abstract}

\vspace{-0.02in}

\vspace{-0.05in}

\section{Introduction} 
\label{sec:introduction}
\begin{figure*}[!h]
	\centering
	\includegraphics[width=0.7\linewidth,keepaspectratio]{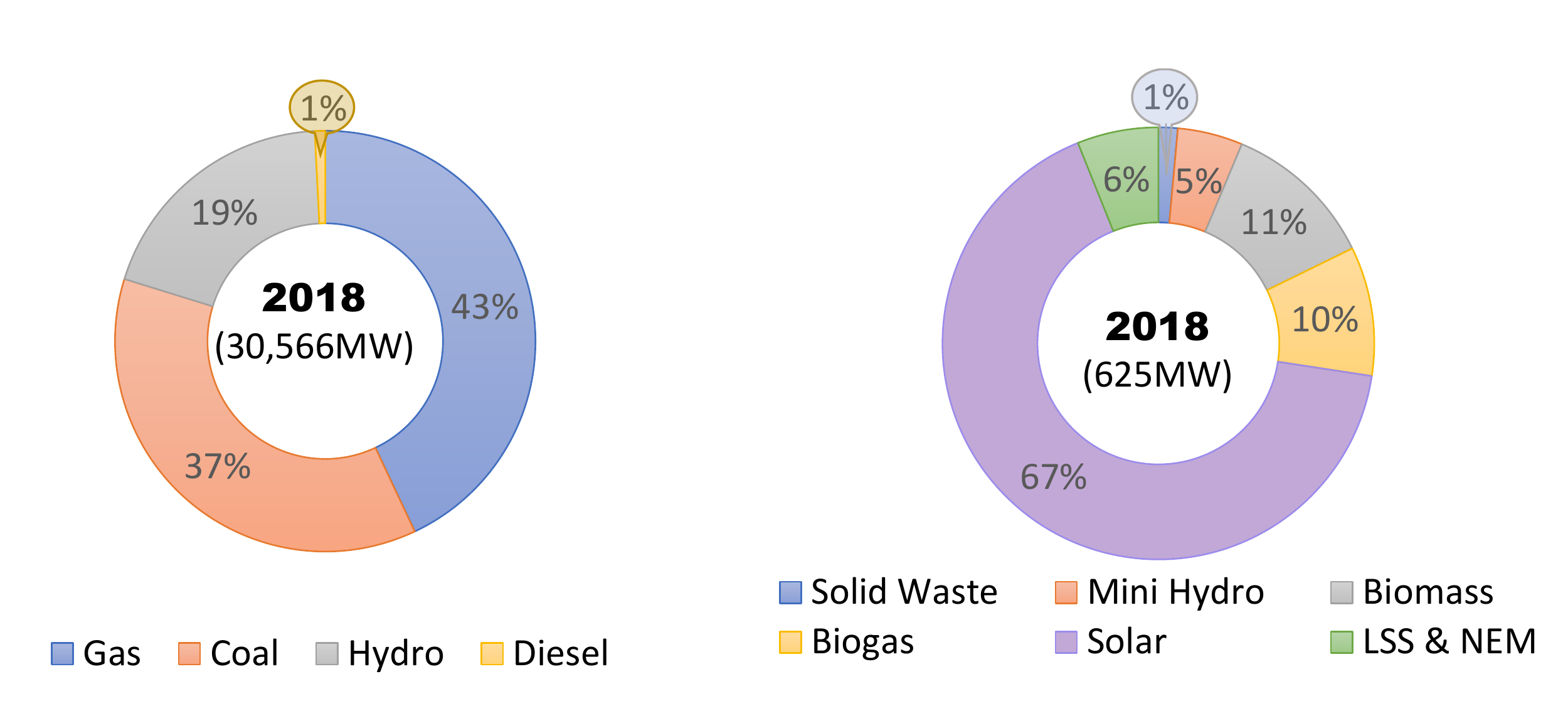}
\vspace{-0.1in}
\caption{Installed energy capacity in Malaysia by plant type (end of 2018). (a) non-renewable sources. (b) renewable sources. LSS: large scale solar, NEM: net energy metering.}
\label{Fig:Fig1}
\vspace{-0.1in}
\end{figure*}

The increased demand for energy across the globe has led to Malaysia recognizing the significance of renewable energy as a supplement to conventional sources in generating electricity.  Malaysia has established several energy policies that made the country one of the leading in energy production in South East Asia. The first National Energy Policy was coined in 1979 and 2011, the Renewable Energy Act was put into existence. The policy framework has the objectives to identify the energy needs properly, conserving resources and environment, promote sustainable development, and low carbon technology \cite{Oh2018, Shaikh2017}. Malaysia pledged to the United Nations Climate Change Conference 2009 (COP15) to decrease the carbon emissions by 40\% \cite{CommissionEnergy2017} and committed to reducing its Green House Gas emissions by 45\% by 2030 with respect to the year 2005 \cite{Kardooni2016,Mah2019}. The new Net Energy Metering (NEM) mechanism updated in 2019, proposes equal tariff for buying and selling electricity for NEM members. This recent process is expected to engage more customers to adopt renewable technology for example wind turbines and solar panels; however, the current NEM covers only the energy generated from solar panels \cite{Abdullah2019}.  Currently, renewable energy in Malaysia are mainly comes from biomass, biogas, solar, wind, mini hydro \cite{Abdullah2019}.  Figure~\ref{Fig:Fig1} summarizes the installed electricity supply capacity mix for the entire Malaysia (Peninsular Malaysia, Sabah, and Sarawak) as of December 2018 \cite{MalaysianEnergyCommission2019, MalaysianEnergyCommission2019a, SarawakEnergy2018}. Figure~\ref{Fig:Fig1}(a) shows the total installed non-renewable sources dominated by fossil fuels, gas (43\%) and coal (37\%) with hydro plants larger than 100 MW (19\%) and 1\% of diesel generators. On the other hand, the renewable sources (grid-tied only), as shown in  Figure~\ref{Fig:Fig1}(b),  covers solar (67\%), biomass (11\%), biogas (10\%), LSS \& NEM (6\%), mini-hydro (5\%), and solid waste (1\%). The renewable capacity provides a total of 625 MW representing 2\% of total energy capacity in Malaysia. However, Malaysian green technology master plan (GTMP) targets to increase the capacity of renewable sources mix to 20\% by 2025. 

\vspace{-0.2in}

Most of the wind power studies in Malaysia focus on the features and characteristics of wind speed. That is due to Malaysia faces challenges in wind energy production as Malaysia is located at 5$^\circ$ on the north side of the equator, also the sea and land air influence the wind circulation system. It was observed that wind changing by area and month and does not blow uniformly \cite{Islam2011}.  As per the statistics, Malaysia is known to have low wind speeds in comparison to countries with high wind speeds such as Denmark and Netherland. However, several locations in Malaysia have good wind upon a particular time in the year, especially during monsoon seasons. Malaysia has two monsoons, the southwest monsoon (May-September) and northeast monsoon (November-March). The average of the wind speed is usually below 3 m/s, which make it less potential for generating countinuous wind energy output, however, the maximum wind speed falls in the range 6-12 m/s indicating the possibility to generate wind power at certain times only \cite{Masseran2012}. With this speed, there have been several attempts to install wind turbines across the country.  A few installed wind turbines as pilot projects were found in the literature. The first wind turbine of 150 kW was installed at Pulau Terumbu Layang-Layang  (Sabah), and the findings confirmed the success of the project \cite{Najid2009}. In contrast, the turbines of 100 kW installed within a hybrid system at Perhentian Island were failed.  The on-site observations noticed that the turbines did not rotate continuously \cite{Albani2017a}.  Besides, an earlier pilot project was implemented in a shrimp farm in Setiu (Terengganu) with a wind turbine of 3.3 kW. The turbine was connected directly to water pump systems and the aeration on the site , which was estimated to cover almost 40\% of the farm's energy consumption with energy cost of 0.98 RM /kWh each year \cite{Ibrahim2014c}.
 
There are few review papers about wind energy in Malaysia that appeared in the literature. However, these reviews are not recent and most of them concentrate on renewable sources in general and do not discuss the literature in depth \cite{Ashnani2014, Azman2011, Foo2015, Hossain2015, Hosseini2014, Ong2011, Shafie2011}. Among the mentioned papers, only a few reviews concentrate on wind energy. However, these reviews investigate specific factors such as wind speed distribution and spatial models \cite{Lawan2013}, particular aspects like political and regularity support \cite{Ho2016}, and technical issues \cite{Didane2016, Palanichamy2015}. None of these reviews conducted a comprehensive review of all related factors to wind potentialities and other wind energy aspects such as wind forecasting, mapping, turbines designs, and optimal sizing of hybrid renewable sources. Therefore, this paper presents a comprehensive review of wind energy research progress in Malaysia extensively. It comprises all potentiality factors, wind predictions, techno-economic, and design, which are further discussed in detail in the following sections.

\begin{figure*}[!h]
	\centering
	\includegraphics[width=0.7\linewidth,keepaspectratio]{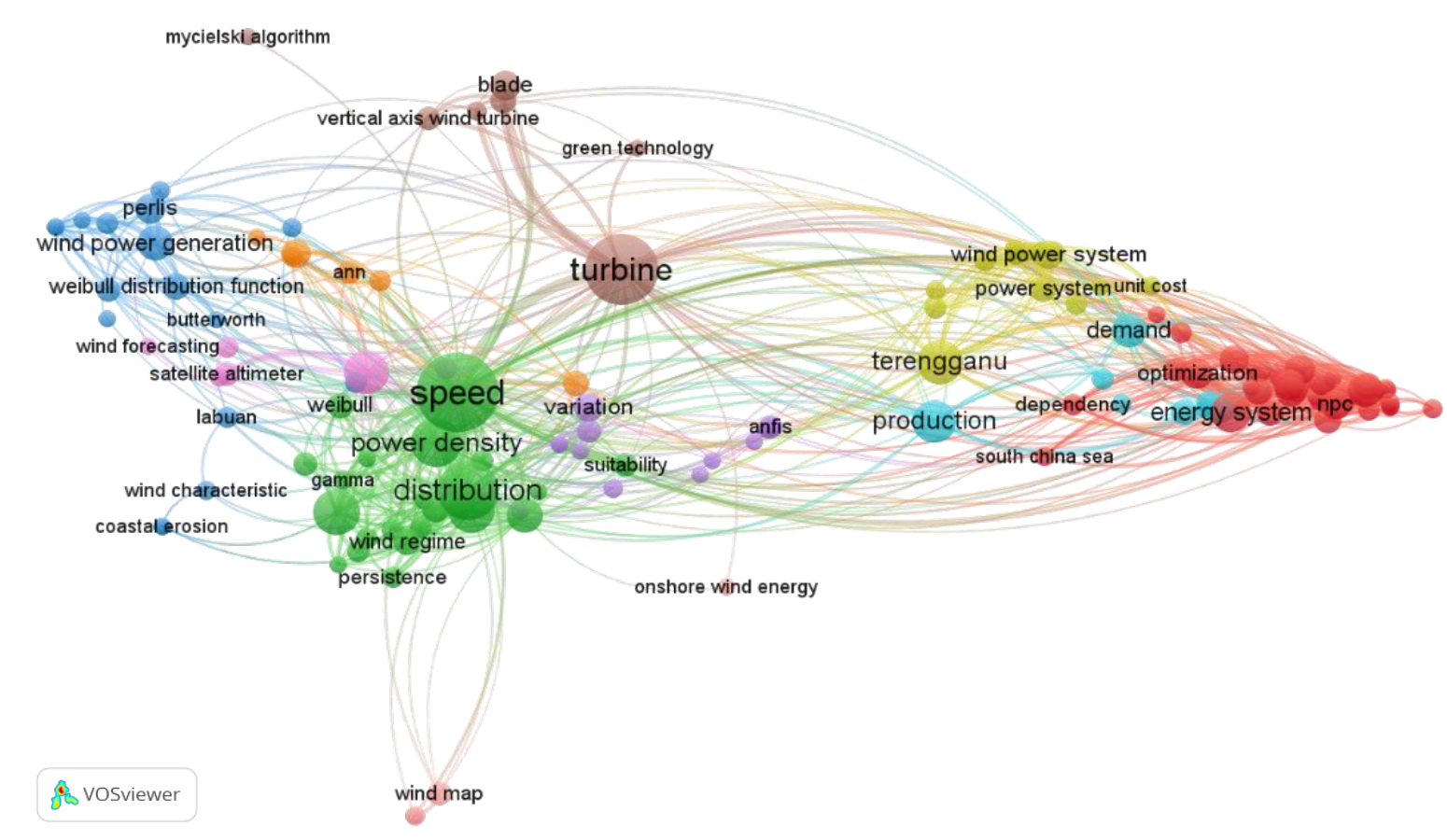}
\caption{The most occurrence keywords in the reviewed articles.}
\label{Fig:Fig2}
\vspace{-0.1in}
\end{figure*}

In this paper, search engines of Sciencce-Direct, Scopus, and Web of Sciences (WoS) were used to find the wind-related studies in Malaysia. From the search queries, a total of 177 articles were found. We first perform titles check, where a total of 27 entries were removed as duplicated articles. The rest of the articles were then subjected to full abstract read. During this stage, a total of 46 articles were removed as out of scope. The remaining 104 entries were entitled for full text read which were categorised as follows: feasibility and potentiality (54); mapping (6); design (7); prediction (8); sizing (18); and reviews (11). Figure~\ref{Fig:Fig2} shows the connections and links of the most occurred keywords in the processed articles in this review. The connectivity map was created using full counting method of all title-abstract keywords \cite{Leydesdorff2016}.

\vspace{-0.1in}
\section{Potentiality}
Learning the wind behaviours is very important to assess the performance of wind farms. The location plays an essential role on wind frequency and speed. Therefore, wind turbines are located only in particular regions globally. To evaluate wind eligibility for the management and design of wind energy system of specific location, various statistical measures and analysis are required to be performed using historical wind data extracted from the intended geographical site \cite{Li2010}. This section reviews energy potential and the statistical evaluation of wind characteristics studies in Malaysia. As shown in Table~\ref{tab1}, the studies categorized based on their objectives. Wind direction, wind speed distribution models, and geographical aspects are presented. Moreover, the Techno-economic effects are reported.

\begin{table}[!t]
\caption{Potentiality studies in Malaysia categorized according to their main objective denoted by * at each column. Geo.: Geographical aspects; WD: wind direction; TE: Techno-economic; Dist.: Distribution modeling; Energy: Energy measure method.}
\label{table}
\centering
\resizebox{0.54\textwidth}{!}{
\begin{tabular}{lccccl}
\hline
Article	&Geo.	&WD 	&TE 	&Dist.	&Energy\\
\hline
\cite{Lawan2015a}      &-&-&-&-&Turbine\\
\cite{irwan2013new}	   &-&-&-&-&electrocorder\\
\cite{Daut2012}	       &-&-&-&*&Power law\\
\cite{Belhamadia2014}  &-&-&-&*&Power law\\
\cite{Sanusi2017} 	   &*&-&-&-&Power law\\
\cite{Anwari2012} 	   &-&*&-&-&HOMOR\\
\cite{Shamshad2007}    &*&-&-&-&turbines\\
\cite{Najid2009}	   &-&-&-&*&NA\\
\cite{Akorede2013} 	   &-&-&*&*&Power law\\
\cite{Yanalagaran2018} &-&*&-&-&NA\\
\cite{Muda2016}		   &-&-&*&-&HOMER\\
\cite{Ngan2012}		   &-&-&*&-&HOMER\\
\cite{Fadaeenejad2014} &-&-&*&-&iHOGA tool\\
\cite{Mekhilef2011}    &-&-&*&-&HOMER - Offshore\\
\cite{Belhamadia2013}  &-&-&-&*&Power law\\
\cite{Albani2013}	   &-&-&-&*&Power law\\
\cite{Islam2011}	   &-&-&-&*&Power law\\
\cite{Irwanto2014}	   &-&-&-&*&Power law\\
\cite{Sanusi2018a}     &-&*&-&*&Power law\\
\cite{Ahmadian2013a}   &-&-&-&*&Power law \& turbines\\
\cite{Sanusi2018}	   &-&-&-&*&Power law\\
\cite{Suboh2019} 	   &-&-&*&-&turbines\\
\cite{najid2011}	   &-&-&-&*&Power law\\
\cite{Masseran2015a}   &-&-&-&*&pdf\\
\cite{Nor2014}         &-&-&*&*&turbines\\
\cite{mekhilef2012}    &-&-&*&-&HOMER - Offshore\\
\cite{Muda2018}        &-&-&*&-&HOMER\\
\cite{Albani2014}      &-&-&*&-&Turbines, WindPRP and WaSP\\
\cite{Albani2017}      &-&-&*&-&NA\\
\cite{Ibrahim2014a}    &-&-&*&-&Turbines - sitting\\
\cite{Masseran2013}    &-&-&-&*&NA\\
\cite{Sopian1995}      &-&-&-&*&Power law\\
\cite{Kadhem2019}      &-&-&-&*&Power law, Sim. turbines\\
\cite{WanNik2019}      &-&-&-&*&pdf\\
\cite{Albani2013}	   &-&-&-&*&Turbines, WindPro, WAsP\\
\cite{Sanusi2013}	   &-&-&-&*&pdf\\
\cite{Albani2017a}	   &-&-&-&*&Sitting, turbines, power law\\
\cite{Lawan2017}	   &-&-&-&*&NA\\
\cite{Ahmed2018}	   &-&-&-&*&NA\\
\cite{Uti2013}		   &*&-&-&-&NA\\
\cite{Lim2019}		   &*&-&-&-&Turbines\\
\cite{Rizeei2018}	   &*&-&-&-&NA\\
\cite{Albani2018}	   &*&-&-&-&power law indexes (PLIs)/ Turbines\\
\cite{Saberi2019}	   &-&-&-&*&Weibull Distribution\\
\cite{Razliana2012}	   &-&-&-&*&Electrocorder\\
\cite{Tiang2012}       &-&*&-&*&Power law \\
\cite{Daut2011}        &-&-&-&*&Weibull Distribution\\
\cite{Syafawati2011}   &-&-&-&*&Weibull Distribution\\
\cite{Masseran2013a}   &-&*&-&-&N/A\\
\cite{Masseran2015a}   &-&*&-&-&Betz's law\\
\cite{Masseran2016}    &-&*&-&-&N/A\\
\cite{Satari2015}      &-&*&-&-&N/A\\
\cite{Izadyar2016}     &*&-&-&-&HOMER\\
\cite{Ahmadian2013}    &-&-&*&-&HOMER\\
\cite{Wahid2019}       &-&-&*&-&HOMER\\
\cite{Arief2019}       &-&-&*&-&HOMER\\
\cite{Shezan2016}      &-&-&*&-&HOMER\\
\cite{Shezan2015}      &-&-&*&-&HOMER\\
\cite{Hossain2017}     &-&-&*&-&HOMER\\
\hline
\end{tabular}}
\label{tab1}
\end{table}

\subsection{Wind Direction }
Shamshad et al., (2007) \cite{Shamshad2007} conducted statistical and power density analysis of wind direction and speed in Mersing. They studied the effects of area topography and roughness on the wind energy using various wind turbines. The results show that the wind direction analysis presents almost 27\% of wind direction occurs in the south-west (240 degrees). However, the analysis of projected wind speed at varying heights indicates that the 240 degrees' direction has the lowest wind speed. 

On the other hand, different distributions models were applied to recognize the best model to determine wind direction in Malaysia. For instance, \cite{Masseran2013a}  found that the finite mixture model of the von Mises (vM) distribution with $H$ amount of components considered the most suitable distribution model. In further investigations, by taking advantage of the flexibility of von Mises distribution model in addressing wind directional data with several modes. Nurulkamal Masseran (2015) \cite{Masseran2015} compared the circular model with the suitability of von Mises model based on nonnegative trigonometric sums. The findings show that the Finite Mixture of von Mises Fisher (mvMF) model with number of $H$ components is $\geq 4$ is the most appropriate model. Furthermore, Nurulkamal Masseran \& Razali, (2016) \cite{Masseran2016} investigated the directions of the wind through the monsoon seasons. They used the data collected from five stations between 1st January 2007 and 30th November 2009 to fit the Finite Mixture model of von Mises distribution to data of wind-direction in hourly basis. The results demonstrated the ability of the model to fit all different data sets with all R2 values $\geq$ 0.9. Similarly, \cite{Satari2015} employed von Mises distribution. They noticed remarkable shifts in the direction of the wind in ten years. However, they stated that the wind direction not impacting the wind speed. 

Uniquely, Sanusi et al., (2018) \cite{Sanusi2018a} used Gama and a finite mixture model of von Mises(vM) distributions for wind speed and direction respectively to assess energy potential in Kuala Terengganu. Analysis of wind power density reveals that the existed uncertainty of using a bivariate model compared to univariate is explaining that the wind with the high speed does not usually concordant with the most common direction of the wind. The study concluded that the combination of Gama and a finite mixture of von Mises distributions was discovered to be the perfect bivariate model to clarify wind data of Kuala Terengganu. Likewise, Sanusi et al., (2017) \cite{Sanusi2017} investigated the effects of wind direction on producing wind energy in Mersing. The finite mixture of the von Mises and Weibull distributions was applied for modelling wind speed and wind direction and speed data respectively. The results show that within the studded time interval (2007-2013), the approximate power density (power equation) ranges from 18.2 to 25 W/m$^2$. Analysis of wind direction also concluded that the south-southwest direction is the dominant wind direction in Mersing.

\subsection{Wind speed distribution models}
Saberi et al., (2019) \cite{Saberi2019} used the Weibull Distribution model to assess the quality of wind data gathered in the year 2017 in Kuala Terengganu. Likewise, Weibull Distribution model has been used as a method to determine the potentiality of wind energy by many studies as shown in Table~\ref{tab2} such as \cite{Daut2011, Deros2017, Rizeei2018, Sanusi2018a, Sanusi2017, Syafawati2011}. It was noticed that the use of Malaysia Metrological Data (MMD) was dominant in most of these studies. The MMD data were collected from stations at 10 m height and mostly, extrapolated to higher heights using the standard 1/7 power law. In Butterworth, analysis of the Weibull distribution function performed using five different wind turbines to calculate the energy density. The calculated yield energy density was 288.23 and 315.10 kWh/m$^2$/year and 315.104 kW/m$^2$/year at hub height of 100 m for 2008 and 2009 respectively \cite{Ahmadian2013}. Likewise, at the height of 100 m and according to the Weibull variables, the mean speed of wind is calculated in hourly basis and then simulated using a joint Sequential Monte Carlo Simulation (SMCS) method and obtained a power densities of 84.59 W/m$^2$, 79.28 W/m$^2$ and 33.36 W/m$^2$ for Mersing, Kudat and Terengganu respectively \cite{Kadhem2019}. For self-collected data,  \cite{Albani2018} gathered wind data at four locations Kudat, Mersing, Kijal, and Langkawi. The authors installed masts then embedded the sensors at different heights. The collected data were used to generate maps of wind resources and analyse power law indexes (PLIs). The results show that the wind speed for a feasible project 5 m/s can be reached only if the tower is greater than 60 meter above ground level in all locations, excluding Kudat, as it is noticed that the average wind speed surpasses 5 m/s at 50 meter above ground level. Furthermore, they assessed the accuracies of exponential fits via matching the findings with the 1/7 law through the capacity factor (CF) discrepancies. As a results Mersing and Kudat show potentiality to adopt medium-scale wind farms, however Kajil and Langkawi are suitable for small-scale wind farms. 

Differently, other studies examined various distribution functions. Tiang \& Ishak (2012) \cite{Tiang2012} employed the function of Rayleigh distribution to investigate wind speed and density in Penang. They found that the annual average of power density is 24.54 W/m2, also predicted the monthly annual mean of wind energy which is at 17.98 kW/m2. In southern Sabah,  Masseran et al., (2013) \cite{Masseran2013} used goodness of fit analysis to indicate Gamma and Burr distributions which most of the analysed stations show a good fit. However, it was noticed that Gamma distribution is the best for all stations. Similarly, \cite{Najid2009} applied three different probability distributions (Weibull, Gamma, burr) to evaluate the potentiality of wind energy in Kuala Terengganu. The study used a self-collected wind speed database recorded daily at 18 m hub-height for two years period (2005-2006) with an average speed of 2.78 and 2.81 m/s respectively. They concluded that among the proposed three distributions, Burr distribution provides the best fit of the analysed wind speed data. In another study,  Masseran et al., (2012) \cite{Masseran2012a} evaluated wind speed persistence (stationarity and variability) in 10 wind stations (Kuala Terengganu ,Alor Setar, Bayan Lepas, Kuantan , Chuping, Ipoh, Kota Bahru, Mersing , Malacca,  and Cameron Highlands) during the period from 2007 to 2009 using data obtained from MMD. Wind speed duration curve (WSDC) was used to generate an accumulative distribution of wind speed through a specific duration. The findings demonstrated that the wind speed form Chuping has the lowest variations and the most persistent regardless of the wind speed values (average of 1.03 m/s). However, considering the wind speed values and energy potential, Mersing was recognized to have the most potential site for energy generation with 18.2\% of three years wind speed exceeds 4 m/sin comparison with other sites. In more investigation about Modelling and statistical evaluation of wind power density,  \cite{Masseran2015a} utilized wind data from Cameron Highlands, Mersing,  Malacca, Sandakan,  Kudat,  and Putrajaya stations obtained from MMD to evaluate several modelling methods including, Gamma,  Weibull, and inverse Gamma density functions. The study proposed Monte Carlo integration method to better estimate the statistical parameters from the power density which was calculated from distribution models. The findings show the efficiency and reliability of the integrated method of Monte Carlo in delivering a good solution for defining the statistical attributes of wind power density. The results show that the mean power per unit-area ranges between 2 and 24 W/m$^2$, and Kudat and Mersing sites have the greatest value compared to other stations. Also, Kudat and Mersing stations have the greatest standard deviation with the range from 4 to 45 W/m$^2$.

\begin{table*}
\caption{Studies applied weibull distribution model.}
\label{table}
\centering
\resizebox{0.8\textwidth}{!}{
\setlength{\tabcolsep}{10pt}
\renewcommand{\arraystretch}{1.2}
\begin{tabular}{p{1cm}|p{5cm}|p{4.7cm}|p{3.5cm}|p{6cm}}
\hline
\rowcolor{Gray}
Study 	&Location 	&Power Density (W/m$^2$)	&Wind Speed (m/s)	&Energy Density \\
\hline
\rowcolor{LightCyan}
\cite{Saberi2019} & Kuala Terengganu & Average in 2017 is 43.42 W/m$^2$ & 2.01 & NA \\
\cite{Sanusi2017}& Mersing & 18.2 -- 25 W/m$^2$ & Average between 2--6	&N/A	\\
\rowcolor{LightCyan}
\cite{Islam2011}& Kudat and Labuan & The highest power density is 67.40 W/m$^2$ Kudat and 50.81 W/m$^2$ in Labuan & At Kudat average was 3.37, 3.36 and 3.00 for the years 2006, 2007 and 2008 respectively. At Labuan were 3.50, 3.81 and 2.67 & The maximum is 590.40 kWh/m2/year in Kudat and 445.12 kWh/m$^2$/year at Labuan \\
\cite{Albani2013} &Langkawi	&N/A	&Mean speeds were 2.00 and 2.50 m/s, respectively, at 10 and 30 m	&The range between 68.6--286.7 kWh/m$^2$/year\\
\rowcolor{LightCyan}
\cite{Irwanto2014} &Perlis (Chuping and Kangar)	&In Chuping result is 2.13 W/m2 and in Kangar is 19.69 W/$^2$ for hub-heights above 50 m &	Average 1.2 \& 2.5	&N/A\\
\cite{Ahmadian2013} &Butterworth	&32.61W/m$^2$
and 35.65W/m$^2$ for years 2008 and 2009 respectively	&Average 2.86 m/s	&288.23 and 315.10 kWh/m$^2$/year at hub height of 100 m\\
\rowcolor{LightCyan}
\cite{Kadhem2019} &Mersing, Kudat and Terengganu	& 84.59, 79.28 and 33.36 W/m2 at a height of 100 m &	Annual Mean is 2.82, 2.45, and 2.03 &Annual 61.754, 57.877, and 24.356 kWh/m$^2$\\
\cite{Shamshad2007} &Mersing	&Average of 44 W/m2	&N/A&	N/A\\
\rowcolor{LightCyan}
\cite{Daut2012} &Perlis	&1.1549	&1.7133 &NA\\
\cite{Belhamadia2014} &Mersing, Kota Bharu, Miri, Kuala Terengganu, and Lingkawi	& Mersing shows the highest 69.257 w/m$^2$ at hub-height of 65 m 	&2.74, 2.27, 2.18, 2.05, and 1.92	&N/A\\
\rowcolor{LightCyan}
\cite{Belhamadia2013} &Kota Bharu, Kuala Terengganu, Langkawi, Mersing, Miri	&Mersing shows the highest 69.257 w/m$^2$ at hub-height of 65m followed Kota Bharu  45.174 w/m$^2$	&Mersing shows the highest of 3.97
	&N/A\\
\cite{Albani2013} &Kudat	&21 W/m$^2$	&2.8 m/s	&14.6 MWh/year\\
\rowcolor{LightCyan}
\cite{Sanusi2018a} &Kuala Terengganu	&The univariate model shows a value of less than 4 W/m$^2$ and mostly higher than 4 W/m$^2$ for the bivariate distribution	&The highest mean is 2.9201 m/s &N/A\\
\cite{Razliana2012} &Kangar, Perlis	&The highest power density is 64.206  W/m$^2$	&Average of 3.5 	&N/A\\
\rowcolor{LightCyan}
\cite{WanNik2019} &Peninsular Malaysia	&The highest and lowest values of the power density of 84.60 and 15.20 W/m$^2$	&The monthly average wind speeds ranges from 2.00 m/s to 5.20 m/s&	N/A\\
\cite{Lawan2017}& Sarawak  	&The highest and lowest values of power density in Marudi are 3.93 and 18.67 W/m$^2$ and in Kula Baram are  3.73 and 18.02 W/m$^2$	&The average annual wind speed is 2.0 m/s &	The highest and lowest values of energy density in Marudi are 34.39 and 163.51 (kWh/m$^2$/year) and in Kula Baram are  32.66 and 158.82 (kWh/m$^2$/year)\\

\hline
\end{tabular}}
\label{tab2}
\end{table*}

\subsection{Geographical aspects }

It was noticeable that many studies were considering some geographical aspects of determining wind potentiality in Malaysia. Lawan et al. (2015) \cite{Lawan2015a} proposed a methodology to evaluate wind energy without using direct wind speed measurements. The study covers some terrain areas in Sarawak (Kuching, Samarahan, Serian, Lundu). A topological feed-forward neural network (T-FFNN) method was used to model the monthly wind speed profile using region-specific geographical, metrological, synthesized topological parameters. The study concluded that the wind power of all areas falls within the lower power density class ($\leq$100 W/m$^2$) with annual energy output in the range of 4--12 MWh/year. In another study, Lawan et al., (2017)\cite{Lawan2017} performed a study of wind power generation at  Miri in Sarawak by analysing the characteristics of wind speed at the height of 10--40 m using the method of ground wind station and T-FFNN. With nine meteorological, geographical and topographical data as input parameters, the model uses monthly wind speed as its output variables. The value of 5800-13,622 kWh/year for the annual energy output (AEO) obtained from their micro sitting analysis shows the possibility of harnessing the wind energy for small scale applications. In Peninsular Malaysia, Yanalagaran \& Ramli, (2018) \cite{Yanalagaran2018} investigated the relationship between the wind flow and coastal erosion over the period from 1984 to 2016 using the image layering technique on wind data acquired from MMD and coastal erosion data collected from newspaper articles, news videos, social media, and other relevant media. Using the equilibrium beach profile theory, the wind and coastal erosion analysis were investigated with the use of wind direction data. The findings demonstrate the good correlation between the wind direction and erosion direction.

A few researchers employed satellite data to examine wind potentiality in offshore and onshore locations. Uti et al., (2013) \cite{Uti2013} used a multi-mission satellite altimeter to study offshore wind potential locations. They extracted the wind speed data from nine different satellites between 1993 and 2016. Meanwhile, they gathered data from buoys at the sea to compare it with satellite data for reliability validation purposes.  The findings of their study indicated that the wind speed measures from the satellites are a little greater in comparison with the wind speed measures from the buoys. However, validation analysis using satellite track analysis and the root mean square error (RMSE) computation proofs a good correlation with positive values 0.6148 and 0.7976 for Sarawak and Sabah respectively. Another study by \cite{Albani2014},  investigated offshore wind resources in Kijal (Terengganu) using QuickSCAT satellite data obtained from WindPRO software in the period from 2000 to 2008 with a horizontal resolution of 12.5km$^2$. WindPRP and WaSP software were used to conduct the potentiality analysis of wind data along South China Sea in Kijal. The optimum Feed in Tariff (FiT) rates as proposed for offshore wind projects in Kijal was found to be about RM 0.81--RM 1.38. The findings concluded that Malaysia has potential for adopting medium and small-scale wind turbines. In onshore areas, Rizeei et al., (2018)\cite{Rizeei2018} used the satellite data to derive solar irradiance estimation, and MMD data to provide daily wind over Peninsular Malaysia. Also, they applied the method of simple additive weighting in the geographic information system (GIS) platform to build the hybrid renewable energy suitability model. The result of the analysis shows that coastal areas at Hulu Terengganu have a good potential for the PV-WT hybrid system. 

\subsection{Techno-economic}

From the techno-economic perspective, many studies have been accomplished by finding the FiT rates to determine the feasibility of wind farms in Malaysia. For instance, Ibrahim \& Albani, (2014) \cite{Ibrahim2014a} selected 22 kW rated power wind turbines for the annual energy production analysis in Kudat. The analysis identified the best site to install wind turbines with a potential production ranging from 37.5--43.1 MWh/Year with FiT rates around RM 0.46-RM 0.80 per kWh. Moreover, Albani et al., (2017) \cite{Albani2017} proposed a method to calculate the optimal FiT suitable for wind energy generation in Malaysia. They selected Kudat as a case study with real data collected at different heights with an average speed of 5 m/s when the hub-height exceeds 30m. The base case FiT results indicate 0.9245--1.1313 RM/kWh for small-scale turbines and between 0.7396 and 0.9050 RM/kWh for utility-scale wind turbines. The analysis of the capacity factor by using different wind turbines was conducted by \cite{Akorede2013}. The findings show that at a height of 36.6 m Mersing achieved the maximum capacity factor (CF) of 4.39\% with wind turbine of capacities 20 kW and 50 kW. Mersing also has the highest energy production of 378 MWh/Year then Chuping with an annual yield of 254 MWh. Mersing achieved the lowest production cost, which about 21 -35 cents followed by Chuping with the highest cost of production at 70 cents for 600 kW turbines. On the other hand, offshore wind resources in Kijal (Terengganu) were assessed by Albani et al., (2014) \cite{Albani2014}. Wind turbine capacity factor analysis shows that the turbine with a rated power of 850 kW was the most suitable for installation in Kijal with capacity factor of 26.8\% (annual energy production of 3.653 MWh with PBP from 7 to 10 years) compared to other evaluated turbines. The analysis shows optimal FiT rates to be around RM 0.81 to RM 1.38. In another offshore site (near to Terengganu coast), a model of 2000 kW wind turbine was taken for gross energy and economic feasibility analysis. The sensitivity analysis of FiT ratio is found to be between 1 and 3 \cite{Mekhilef2011}. Notably, the FiT rate is changed when the level of economic parameters changes.

In another avenue, Izadyar et al., (2016) \cite{Izadyar2016} selected a list of potential locations to set the optimal configuration of the hybrid renewable systems. Other researchers have considered a techno-economic factor in assessing wind potential locations and determining the optimal configurations of hybrid systems \cite{Hossain2017, Muda2018, Shezan2016, Shezan2015, Wahid2019}.  Based on Net Present Cost (NPC), Langkawi was found to be the most potential location for the PV-Wind configuration (NPC of 696,083 USD). Next, Tioman island and Borneo island were found to be fit for the PV-WT- MHP and PV-MHP configuration respectively \cite{Izadyar2016}.  By considering the cost of energy (COE), Ngan \& Tan (2012)\cite{Ngan2012} conducted potential energy and economic analysis on using hybrid PV-WT-DG power with and without a battery storage system in Johor Bahru using HOMER software. Hybrid WT-DG without battery system shows a very high excess electricity from which they finally suggested to add battery storage. The hybrid WT-DG with battery storage shows COE values ranging from RM 1.365/kWh to RM 1.638/kWh. Likewise, Muda Et Al., (2018) \cite{Muda2018} focused on evaluating the wave energy system in term of its feasibility to provide electricity to Terengganu (Pulau Perhentian island) by proposing and analysing hybrid WT-PV-hydro-DG power system. Findings demonstrate that a hybrid wave energy system achieved a lower cost of electricity (COE) which is RM 1.303/kWh in comparison to the hybrid PV-battery system which achieve COE of RM 1.262/kWh. Moreover, energy analysis shows a total of 712,813 kWh/year (81 kWh) of generated power with only a 2.4\% share from wind energy and 0.76\% share from wave energy. In another study conducted in Terengganu, a hybrid system (grid-PV-WT) connected to the grid for households was evaluated. The findings show that the most feasible configuration is the PV connected to the grid hybrid system (2 kW PV panels) with a cost of energy at 0.001 \$/kWh. In other configurations, the total electricity produced by wind turbines is only 14\% and 22\% of total electricity production in PV-WT-grid and wind-grid systems respectively. The analysis shows that the best configuration is PV-WT-grid with an average wind speed of 3.43 m/s and FiT rate of 0.5169 \$/kWh \cite{Muda2016}. In contrast, Anwari et al., (2012)\cite{Anwari2012} studied the case of Pemanggil Island (near Mersing) that has annual wind speed around 1.7 to 6.7 m/s. They found that the (WT-DG) hybrid system can be feasible in case that the wind speed is above 3 m/s and the diesel price is greater than \$ 1.02/Litter.

Elsewhere, at five locations in Malaysia (Kota Belud, Kudat, Langkawi, Gebeng, Kerteh) Nor et al., (2014)\cite{Nor2014} conducted the economic-feasibility analysis of installing 77 Leitwind wind turbines of 1,000 kW each . The study shows that the Kota Belud achieves the highest internal rate of return of 21\% and the lowest payback period of 4.25 years \cite{Nor2014}. In contrast, Suboh et al. ( 2019) \cite{Suboh2019} proposed a dispatch strategy micro-wind turbine farm with storage using monthly average wind data for Mersing in 2009. The study considered a 10 $\times$ 300 W wind turbines (Infiniti 300) with the cut-in speed of 2 m/s. The study concluded that, according to the proposed strategy, the payback period requires 20 years to reimburse capital expenditure. As shown in Table~\ref{tab3} the PV-WT-DG configuration is the optimal setting for Berjaya, Kuala lumper International Airport KLIA Sepang, and Johor. However, few studies found wind energy not feasible to be utilized in several locations in Sarawak and Terengganu for large-scale turbines as the average wind speed falls below 4 m/s \cite{Arief2019, Ahmadian2013a}.

\begin{table}
\caption{Hybrids renewable systems.}
\label{table}
\centering
\setlength{\tabcolsep}{3pt}
\begin{tabular}{p{2cm}|m{4cm}|p{4cm}|p{4cm}}
\hline
\rowcolor{Gray}
Article	& Location	& Optimal Configuration	& COE-Cost \\
\hline
\cite{Muda2018} &Pulau Perhentian Kecil	&PV-DG	&RM 1.262/kWh\\
\rowcolor{LightCyan}
\cite{Izadyar2016}&Langkawi	&PV-WT	& NA\\
\cite{Izadyar2016}&Tioman island	&PV-WT-MHP	&NA\\
\rowcolor{LightCyan}
\cite{Wahid2019}&Kerteh	&PV-WT	&USD 0.474/kWh\\ 
\cite{Shezan2016}&KLIA Sepang (Selangor)	&PV-WT-DG	&USD 0.625/kWh\\
\rowcolor{LightCyan}
\cite{Shezan2015}&KLIA Sepang (Selangor)	&PV-WT-DG	&USD 5.126 /kWh\\
\cite{Hossain2017}&Berjaya Tioman	&PV-WT-DG	&USD 0.279 /kWh\\
\rowcolor{LightCyan}
\cite{Ngan2012}&Johor	&PV-WT-DG	&RM 1.365--1.638/kWh\\

\hline
\end{tabular}
\vspace{-0.2in}
\label{tab3}
\end{table}

\vspace{-0.1in}
\section{Wind prediction }
Optimal prediction of wind speed is an open problem of research due to the need to foresee the feasibility of harnessing wind energy form specific sites. Wind speed forecasting has been extensively studied with hundreds of articles found in the literature. However, wind speed prediction methods can be categorized into five groups: persistent method, physical models, statistical models, artificial intelligence methods, and hybrid methods \cite{Khodayar2017}. A breif summary of these methods is as follows,

\begin{enumerate}
\item The persistent method is the most common baseline method which often used as a benchmark method for comparisons with other methods. It assumes that the wind speed will remain constant over time such that the wind speed value in the future is similar to the current value when the forecast is made. However, due to the intermittent nature of wind speed, this assumption fails when the prediction horizon increases.

\item The physical model is based on weather forecast data (e.g. temperature, pressure, surface roughness, and obstacles) and numerical weather prediction (NWP) models. NWP utilizes complex mathematical models to create a terrain-specific weather condition which are then used for wind speed predictions. NWP requires large computational resources with high time consumption; hence, it is not suitable for short-term wind speed predictions \cite{Eze2019, Khodayar2017}. 

\item The statistical methods learn the correlations of historical wind speed data and/or other explanatory variables (i.e. the common metrological parameters or NWPs). Statistical models are usually less complex than the physical models and show accurate performance when used for short-term forecasting tasks. The conventional statistical techniques can be categorized into linear stationary models, nonlinear stationary models, and linear non-stationary models. Generally, the statistical models are limited by data underlying dynamics constrained by a strict assumption of normality, linearity or variable independence \cite{Mishra2018}.

\item The artificial intelligence methods and supervised mechanisms are the dominant fields of research in recent years which can represent any complex relationship between the data regardless of the underlying dynamics  \cite{Chang2014}. Recent advances in deep learning methods have seen significant performance improvements in many practical forecasting tasks, sometimes surpassing the known strong statistical models' performance. This is owing to the capability of deep learning methods to perform both feature extraction and modeling which allows learning of complex data representations with the privilege of hierarchical levels of semantic abstraction via multiple stacked hidden layers and hence the robust and accurate forecasting even based on raw data from multiple exogenous or heterogeneous sources.

\item Hybrid models are based on the combination or integration of two or more models of different methodologies to benefit from the advantages of each model to improve prediction performance \cite{Khodayar2017, Chang2014}.
\end{enumerate}

In this paper, we limit the review of wind speed forecasting methods to those studies that have been carried out in Malaysia using data collected in Malaysia too. A limited number of research articles found in the literature which tried to improve the prediction performance mostly using metrological station data. We conduct a thorough review of eight published articles related to wind data prediction in Malaysia during the period from 2005 to 2018 in the analysed database. Four prediction methods were used artificial neural network (ANN) \cite{Kadhem2017, Lawan2015, Shamshad2005, Shukur2015},  two deployed adaptive neuro-fuzzy inference system (ANFIS) \cite{Hossain2018, Sarkar2019}, and two used seasonal autoregressive integrated moving average (SARIMA) \cite{Deros2017}, and Mycielski algorithm \cite{Goh2016}. 

The first article of wind speed prediction for Malaysia was published by Shamshad et al. (2005) \cite{Shamshad2005}. The authors implemented a stochastic synthetic generation of wind data using Markov chain with wind data obtained from the MMD for two stations (Mersing and Kuantan) during the period from 1995 to 2001. A first and second-order matrix of the Markov process with 12 states were estimated from the wind speed data which are then reversely used to predict future wind speed data. To assess the performance of Markov chain matrices, the Weibull distribution was estimated from observed and predicted wind data. Both transition matrices show comparable distribution parameters of k and c. Autocorrelation and spectral power density were also used for performance evaluation when using different correlation time lags. The RMSE of the first and second-order was 0.12 and 0.03 for Mersing and 0.03 and 0.03 for Kuantan. Markov model is one type of statistical model that heavily relies on probabilistic theory and higher orders show better performance with more state-space complexity. 

 Lawan et al. (2015) \cite{Lawan2015} proposed a topological ANN method for monthly wind speed prediction. The geographical and metrological data are obtained from MMD for eight regions in Sarawak for a period of ten years from 2003--2012 (Kuching, Miri, Sibu, Bintulu, and Sri Aman) and five years from 2008 to 2012 (Kapit, Limbang, and Mulu). Nine parameters are used as input to the ANN model including Latitude, Longitude, Altitude, Terrain Elevation, Terrain Roughness, month, temperature, atmospheric pressure, and relative humidity. For each station, the data were split into 70\%, 20\%, 10\% for train, test, and validation respectively. All data were normalized to the interval of [-1, 1]. Monthly prediction results show the minimum and maximum values of correlation coefficient $R$ were 0.8416--0.9120, and the maximum MAPE of 6.46\%. 

Another study by Shukur \& Lee, (2015) \cite{Shukur2015} suggested a hybrid method of ANN and Kalman filter (KF) to improve the performance of ANN models. However, the authors used daily wind speed data obtained from MMD for Muar station during the period from 2006 to 2010. Unlike the previous studies which mostly use data-driven prediction models, state-space model parameters were estimated using the ARIMA model and then used as input to ANN to predict the wind speed future data. ARIMA (0,1,2) (0,1,1)$_12$ model was selected for state-space parameters estimation with MAPE of 19.27, the Hybrid ARIMA-KF achieved 17.20 MAPE, and Hybrid KF-ANN showed 11.29 MAPE. The results indicate that KF-ANN improved the prediction performance by almost 8\% compared to the direct use of the ARIMA model. However, the state-space models are usually of high computational complexity and sensitive to the dynamics of data, especially that of the high sampling rate. 

Kadhem et al. (2017) \cite{Kadhem2017} proposed a short-term wind speed prediction hybrid model using Weibull distribution and ANN (HANN). Hourly wind speed data were obtained from MDD from three stations (Mersing, Kudat, and Kuala Terengganu) for three years 2013--2015. Data were normalized in the interval [-1, 1] and then partitioned into 60\% train, 20\% validation, and 20\% test. Five parameters are used as input to the ANN model including wind speed of Weibull model, direction, hour, day, month. Results show the proposed HANN method achieved a MAPE and RMSE of 6.06\% and 0.048 respectively for Mersing data which improves the prediction performance for 10.34\% MAPE compared to the ANN model alone. 

The suitability of using a hybrid ANFIS system and optimization methods for long-term (weekly and monthly) wind prediction in Malaysia was investigated by Hossain et al. (2018) \cite{Hossain2018}. Wind speed data were obtained from MMD of several locations Mersing, Kuala Terengganu, Pulau Langkawi and Bayan Lepas during the period 2004 to 2014. The data were adjusted to 50m hub-height. Three optimization techniques namely, particle swarm optimization (PSO), genetic algorithm (GA), and differential evaluation (DE) were used to optimize the ANFIS membership function parameters. Method performance was analyzed when using different train-test data partitioning [70\%,30\%], [60\%,40\%], and [80\%,20\%]. Both ANFIS-PSO and ANFIS-GA seem to be able to provide an overall good power density prediction performance. The proposed hybrid ANFIS model was also used for daily wind data extrapolation of Tioman island where no MMD station is available. Even though ANFIS is a combination of ANN and fuzzy inference system (FIS) which supposed to overcome the limitations of both methods and provide better performance.

Sahin \& Erol, (2017) \cite{Sahin2017} conducted a comparative analysis of both ANN and ANFIS concluding that the ANN approach performed better than the ANFIS model on predictions. Sarkar et al. (2019) \cite{Sarkar2019} also conducted a comparative study of wind power prediction using ANFIS model in Kuala Lumpur and Melaka. Wind data were obtained from MMD over the period 2013 to 2015. Four input variables are used for prediction including humidity, pressure, wind speed, and temperature. With one year of test-data, results show satisfactory results when using ANFIS model to predict wind power data of Kuala Lumpur compared to Melaka with (RMSE=1.82, MAE=1.98, MAPE=13.23 and R2=0.944) for Melaka and (RMSE=1.65, MAE=1.87, MAPE=12.35 and R2 = 0.968) for Kuala Lumpur. 

Mycielski algorithm is a simple non-parametric alternative approach to the complex parametric models which is mainly based on pattern matching \cite{Croonenbroeck2015}. Goh et al. (2016) \cite{Goh2016} proposed a wind energy assessment comparing two methods (Mycielski algorithm and K-means clustering) for wind prediction in Kudat, Malaysia. Wind data were obtained from MMD over the period 2002 to 2010, the data from 2002--2009 were used for training while 2010 for testing. Results show that K-means clustering provides more accurate predictions compared to the Mycielski algorithm. Weibull distribution was used to assess the comparison of predicted wind speeds with actual ones with prediction RMSE of Mycielski was 1.875 and 1.391 for K-means. 

Wind profile in Malaysia is highly dependent on monsoon seasons, where four main seasons take turns throughout the year producing different wind speed and direction behaviours. Deros et al. (2017) \cite{Deros2017} proposed a wind speed seasonal forecasting model using SARIMA with wind data obtained from MMD during the period 2000 to 2015 for Langkawi. Four monsoons were analysed including Northeast Monsoon, April Inter-monsoon, Southwest Monsoon, and October Inter-monsoon. For variations analysis and determination of seasonal wind distribution, three different distribution models were investigated including Gamma, Lognormal and Weibull distribution. Forecasting results show that SARIMA achieves RMSE=0.3186, MAE=0.265, and MAPE=11.644\%. Among the three tested distribution models, the Log-normal distribution model produced the lowest gap value among all the seasonal wind speed data in each monsoon season. The study concluded that the northeast (NE) monsoon acquires the highest mean wind speed with an average between 1.8 to 2.3 m/s, followed by October inter-monsoon, Southwest monsoon and the lowest wind speed was estimated to be in April inter-monsoon between 0.9 to 1.3 m/s.

According to the reviewed articles on wind speed prediction, we noticed that the lack of standardization between the proposed methods and wind database makes it difficult for technical or numerical performance comparisons. It's important to understand the performance of these models and provide some measure of confidence. Fare analysis requires the prediction experiments to be conducted under similar environments and conditions, such as data sampling rate, train-test partition, prediction horizons, input variables, target area, and hub-heights. Hence, a benchmark study is required to test the generalization, universality, and transferability of these prediction models.

\vspace{-0.1in}
\section{Turbines design }
In the continuous search for the possibility of generating power from wind energy in a low wind speed region, a considerable amount of research works has been reported in that respect. The most prevailing is to design wind turbines specifically for low wind speed region. (Wen et al., 2019a) \cite{Wen2019a} applied blade element momentum theory (BEMT) to design and study the performance analysis of six constant speed horizontal axis wind turbines (CS-HAWT). On the bases of Airfoils low Reynolds number that will fit small wind turbines blade design for low wind speed region application, Six Airfoils: BW-3, FX63-137, NACA4412, S822, SG6040 and SG6043 with a Reynolds number ($<$ 500,000) characteristics and a tip speed ratio of 5 and 6 were chosen for the design. Their results showed that by increasing the Reynold's number the aerodynamic performance of the Airfoils will rapidly increase as well and will, in turn result in lower cut-in speed as well as a higher power coefficient. A blade of 1.78 m/s cut-in wind speed and a higher power coefficient of 0.52 is obtained at the TSR of 5 and 1.8m/s cut-in wind speed, 0.51 power coefficient at the TSR of 6 with Airfoils SG6043 and FX63-137. A lower cut-in speed of 1.5m/s was obtained with Airfoils BW-3 and FX63-137. The lowest cut-in speed was recorded at the TSR of 5 with the Airfoils. Their results of computed annual energy generation (AEG) when the designed turbine power curve are employed with the wind speed distribution of Kangar, shows that out of the six airfoiled turbine selected, SG6043 has outperformed all with AEG of 64kWh and 66kWh for the 5 and 6 TSR respectively. 

Similar research was conducted by Wen et al., (2019b) \cite{Wen2019} to design a small wind turbine's blade as partial solution to the low wind speed region of Malaysia. Based on Malaysia's average wind speed of 2-3 m/s, they chose NACA 4412 Airfoilss profile with lower Reynold numbers of Re0.2, Re0.25, Re0.3, Re0.35 and a tip speed ratio (TSR) of 5, 6 and 7 to design a small wind turbines of 4 m/s rated wind speed. A turbine blade that starts rotating at a wind speed of 1.5 m/s and 2.0 m/s with the efficiency of about 49.31\% was realized. Their result of the AEG using the designed turbine's power curve and the Weibull distribution function  with wind profiles of Kuching, Miri, Kangar, Labuan and Kudat shows a better performance of SWT\_re0.25 and SWT\_re0.3. Misaran et al. (2017) \cite{Misaran2017} designed and developed a hybrid vertical axis wind turbine that can start rotating at the wind speed of 1.0 m/s and can generate a useful power from a low and intermittently high wind speed region like Malaysia. The blade designed was carryout with a CAD software and utilizes airfoils tools 14 to generate the NACA0012 profile.

In another perspective Johari et al., (2018) \cite{Johari2018} used CATIA software to design and compare the performance of three-blade horizontal axis wind turbines (HAWT) and Darrius type vertical axis wind turbine (VAWT). The result of the HAWT and VAWT performance analysis test shows that the HAWT produced higher voltage at stable wind condition but the voltage dies up as the wind direction changes. As for the VATW even though its output voltage is not as high as that of HAWT but it remained unaffected by the change in the wind direction.  Relatively, a vertical axis wind turbine's blade based on Aeolos-V1k was modified and redesigned using ANSYS software by \cite{Yi2018}. The modification which is performed with teak wood material reduces the blade's weight and cost and improves its efficiency.

Several articles to further explore the possibility of generating electricity from the wind energy sources in Malaysia have been reported. Awal et al. (2015) \cite{Awal2015} designed a vehicle-mounted wind turbine (VMWT) to utilize the speed of the wind that passes the vehicle as it moves and to generates electricity. It uses a permanent magnet DC motor as a generator (PMDC) and produces 150-200W of power as the vehicle moves. Likewise, Khai et al. (2018) \cite{Khai2018} modified the rooftop ventilator to functions as wind turbine and Harvest wind energy from the steady wind speed of the ventilation channel that is created due to the temperature difference between the indoor and outdoor. The result of their testing with a stand fan shows that the modified wind turbine can produce a minimum of 4.63 V and 21.44 mW power which can light up a 5 V LED bulb. Moreover, the design and implementation of a VAWT that utilizes the waste exhaust coming out from the cooling tower was presented by Rahman et al. (2015). It comprises of the drag blade (C-type), a guide-vane to increase the inlet wind speed and an enclosure to avoid the wind coming from the opposite direction. They have recorded a voltage and a current of 1.97 V and 0.0041 A and average power of 0.0058 W and concluded that their system is capable of generating 0.0081 W of electricity. 

\vspace{-0.1in}
\section{Optimal sizing of hybrid energy systems }
\label{sec:sizing}
The integration of conventional energy sources with renewable energy sources and energy storage devices is common in developing hybrid energy systems to fulfill a particular load demand. However, the involvement energy storage system is essential to moderate the supply-demand mismatch. Moreover, conventional sources, like diesel generators or fuel cells can also be added to the renewable sources to get a better energy balance.  The hybrid systems begin to be common nowadays as they merge the advantages of both AC and DC- as well as flexibility, cost-effectiveness and to integrate loads and sources based on their characteristics \cite{Hosseinalizadeh2016, Sanajaoba2016}. However, there is no one configuration can fit all solutions, therefore, the most suitable configuration must be considered for a particular application and site. 

The literature of wind energy in Malaysia shows two software tools that have been used in optimizing hybrid renewable energy systems. The first software tool is HOMER that has been used mostly as shown in Table~\ref{tab4}, and the second is iHOGA that was applied by Fadaeenejad et al. (2014)\cite{Fadaeenejad2014}. In contrast, other studies \cite{Khatib2015, Khatib2012, DeAn2011} used MATLAB simulation. Energy systems connected to the grid or in standalone mode involving conventional renewable sources and storage systems can be optimized and simulated using HOMER. The optimization can be executed using historic meteorological data of the location.  The HOMER software was used to find the optimal configuration of hybrid energy systems for different locations in Malaysia. Researchers' main objective was to reduce NPC subjecting it to a few constraints such as reliability and environmental. Table~\ref{tab4} lists the studies from different locations in Malaysia that used the HOMER software tool for size optimization for hybrid energy systems.

Majority of the studies found that PV-WT-DG energy resources are feasible as it provides the minimum NPC and COE with lower CO2 emissions \cite{Haidar2011, Hossain2017, Khatib2012, Mohamed2013, Ngan2012, Shezan2016}. Nevertheless, Khatib et al., (2013) \cite{Khatib2013} examined the power sources in nine locations in Malaysia to determine the cost of energy for every power system as a standalone system. They found that the average cost of the wind energy system is 1.6--7.29 USD/kWh. In comparison, the solar, diesel generator and grid power average cost are 0.35--0.5 USD/kWh, 0.27-0.30 USD/kWh and 0.11 USD/kWh respectively. The findings concluded that using wind energy as a standalone system is not feasible in Malaysia. In concurrent with the previous study, Haidar et al. (2011) \cite{Haidar2011}, used HOMER software to study the feasibility of wind energy within a hybrid system in four locations; Pinang, Johor Baharu, Sarawak, and Selangor. The results show that the cost of wind energy is 1.054--1.457 USD/kWh. The findings lead to the fact that PV-DG is the optimal configuration in Malaysia.

\begin{table*}
\caption{Optimal sizing by using the HOMER software tool.}
\centering
\resizebox{0.95\textwidth}{!}{
\setlength{\tabcolsep}{5pt}
\renewcommand{\arraystretch}{1.2}
\begin{tabular}{l|p{3cm}|c|c|c|c|p{4cm}}
\hline
\rowcolor{Gray}
Study	&Location	&Energy Sources	&Energy storage	&Energy conversion&	Factors	&Key factors \\
\hline
\rowcolor{LightCyan}
\cite{Muda2018}& Pulau Perhentian Kecil	&PV-DG	&Batteries	& CO	& NPC, COE &	Second best combination is PV-WG-DG\\
\cite{Izadyar2016}& Malaysia	&PV-WT-MHP	&Batteries	&CO	&NPC	&Tioman can be considered as the most attractive location for PV-WT- MHP\\
\rowcolor{LightCyan}
\cite{Wahid2019}& Pontian, Kerteh and Teluk	&BG-PV-WT&	Batteries	&CO	&COE&NA	\\
\cite{Shezan2016}& KLIA Sepang (Selangor)	&PV-WT-DG&	Batteries&	CO	&NPC, COE, CO2&NA\\
\rowcolor{LightCyan}
\cite{Hossain2017}& Berjaya Tioman	&PV-WT-DG	&Batteries	CO	&NPC, COE, CO2&NA\\
\cite{Ngan2012}& Johor	&PV-WT-DG	&Batteries	&CO	&NPC, COE, CO2&NA\\
\rowcolor{LightCyan}
\cite{Haidar2011}& Johor, Sarawak, Penang and Selangor&	PV-WT-DG	&Batteries	&CO	&NPC, COE, CO2&	PV-WT proved to be cheaper and more environ- mentally friendly than DG\\
\cite{Ashourian2013}& Tioman Island	&PV-WT-FC	&Batteries&	CO-El	NPC, COE, CO2&NA	\\
\rowcolor{LightCyan}
\cite{Khatib2012}& Bintulu, Kota Kinabalu, Kuala Terengganu, Kuching, Kudat, Mersing, Sandakan, Tawau and Pulau Langkawi&	PV, WT, DG&	Batteries&	CO	&COE&NA\\
\cite{Alsharif2016} & East coast of Peninsular Malaysia&	PV,WT, EG&	Batteries	&CO	&NPC, COE, CO2&	NA\\
\rowcolor{LightCyan}
\cite{Mohamed2013}& Kuala Terengganu 	&PV,WT, DG&	Batteries	&CO	&COE, FiT&NA\\
\hline
\end{tabular}}
\vspace{-0.2in}
\label{tab4}
\end{table*}

\vspace{-0.1in}
\section{Wind mapping }
A comprehensive onshore or offshore wind resources mapping is necessary for assessing wind feasibility especially for Malaysia with a low wind profile. It is well known that wind speed varies according to terrain conditions, such as temperature, wind direction, and height from the ground, surface roughness, day time, and year seasons \cite{Laban2019}. Wind resource geographical maps can be generated by using either spatial prediction or altimetry wind data for the unobserved or uncovered locations by metrological stations. Development of wind energy in Malaysia is still in its early stages, with majority studies use secondary metrological wind data which were collected by weather stations mainly located in airports \cite{Ibrahim2014b}. These wind data were obtained from MMD, which were mainly used in airports for the weather forecasting. Hence, few studies highlighted this issue and conducted nationwide wind energy planning by using spatial interpolation methods to create wind maps for Malaysia \cite{Ibrahim2015}.

Masseran et al., (2012) \cite{Masseran2012} conducted a study to generate wind resources map for entire Malaysia using statistical and geostatistical analysis including semivariogram and inverse distance weighted (IDW) methods for spatial correlation, estimation, and interpolation. A total of 10 years of wind speed data from 2000 to 2009 for 67 stations were obtained from MMD to create a nationwide spatial map. The study first determines the best suitable wind speed distribution. Nine types of wind speed statistical distributions are examined, namely Weibull, Burr, Gamma, Inverse Gamma, Inverse Gaussian, Exponential, Rayleigh, Lognormal, and Erlang. Goodness-of-fit tests indicate only five distributions can be used for each station, Gamma, Burr, Weibull, Erlang and Inverse Gamma to derive the power distribution for each station. Raw moment and Monte Carlo approach were used to determine the mean power. The spatial correlations and dependencies were investigated using the semivariogram method, while the IDW method was used to estimate/interpolate the wind power in the spatial dimension for mapping. Finally, the spatial maps were created. Spatial Mappings of mean power density reveals that the northeast, northwest, and southeast of peninsular Malaysia and southern region of Sabah are the most potential locations for wind energy development.

In 2014, another wind mapping study was conducted by \cite{Ibrahim2014b} covering only nine sites (Mersing, Kuala Terengganu, Pulau Langkawi, Sandakan, Kudat, Kota Kinabalu, Bintulu, Kuching, and Tawau) with only one-year wind data obtained from MMD for 2009. Weibull distribution and IDW interpolation methods were used for mapping. The study concluded that Mersing and Kudat are considered the best potential sites showing an average of 3 m/s above 60 m high. The IDW spatial wind mapping method showed that the area located at the southern part of Peninsular Malaysia has better wind resources. These conclusions were proved by Weibull distributions of each area. The study conducted further analysis of wind speed distribution and direction. However, it has less spatial analysis in terms of data and locations compared to the previous study of Masseran et al. (2012) \cite{Masseran2012}. Same authors in cite{Ibrahim2014} studied the spatial interpolation and mapping of wind distribution in Malaysia with wind data obtained from MMD considering more sites of 12 stations (Langkawi, Kuala Terengganu, Mersing, Kuching, Bintulu, Kota Kinabalu, Kudat, Sandakan, Tawau, Cameron Highland, Kuantan, and Kapit). 

For mapping, the wind extrapolation, power-law, and the Digital Elevation Model (DEM) were together used to create the onshore spatial wind distribution for Malaysia. Then the Ordinary Kriging method was used for spatial wind data were interpolation which is then combined with DEM maps via the power law to produce the final wind map. Results show monthly maps variations with an accuracy of the Kriging interpolation and wind map of $\pm$1.1143 and $\pm$1.4949 m/s respectively. The study was limited to the selection of few sites for mapping, the wind direction was excluded in the analysis, and the produced wind map was limited to 10 m height only. The authors published another article in \cite{Ibrahim2015} following similar methodology and wind data of their previous study in \cite{Ibrahim2014b}. Power density Spatio-temporal maps were generated with extrapolated wind data to different heights from 10 m to 100 m. The study concluded that the wind power of Malaysia is falling into Class 1, showing 1.5232 MW/m$^2$ at 100 m height.
\vspace{-0.2in}
\begin{figure*}[!h]
	\centering
	\includegraphics[width=0.75\linewidth,keepaspectratio]{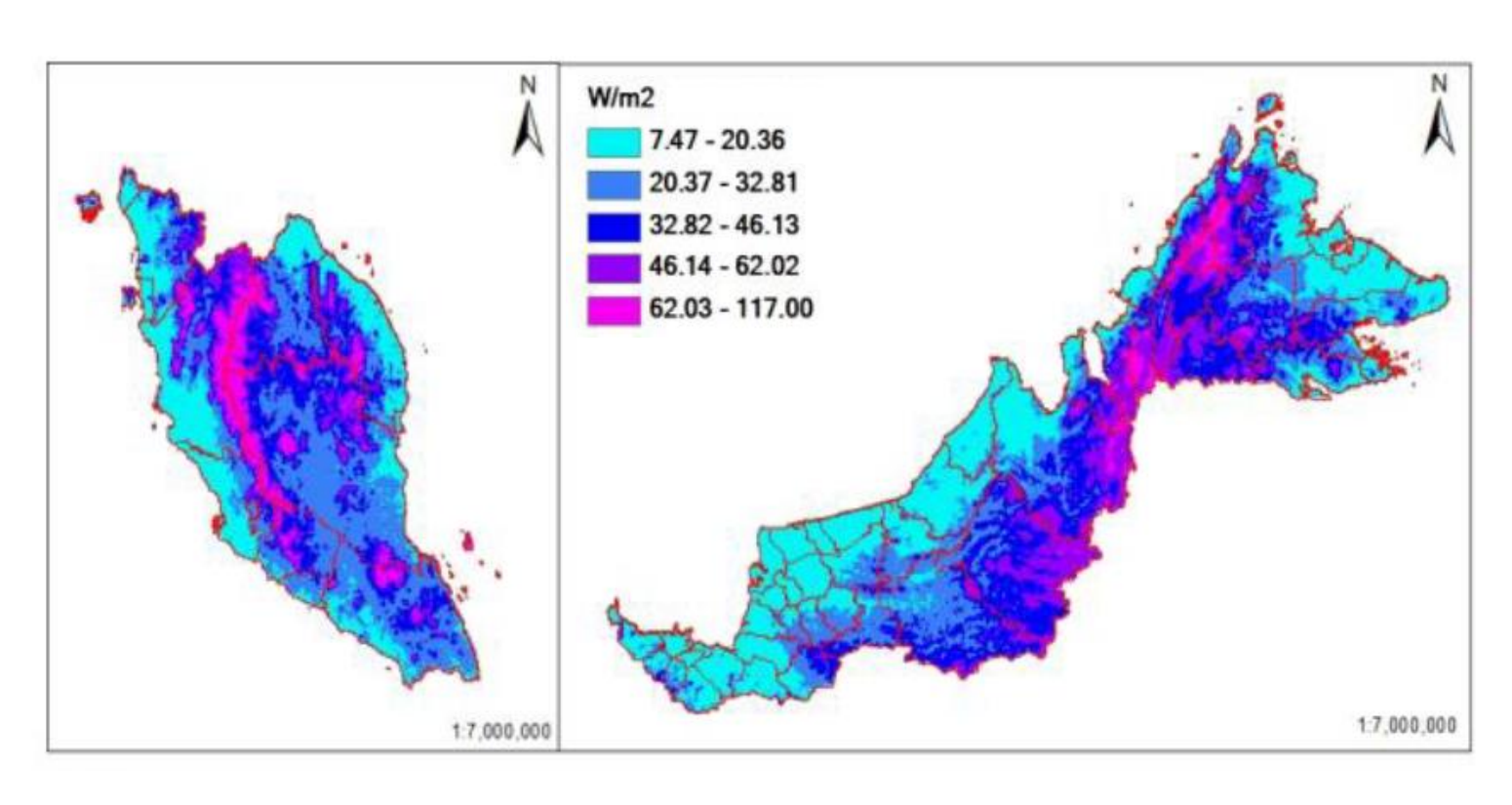}
\vspace{-0.25in}
\caption{Malaysian onshore wind power density map at 10 m height \cite{Ibrahim2015}.}
\label{Fig:Fig3}
\vspace{-0.05in}
\end{figure*}

Partial wind map of the eastern part of Malaysia (i.e. Sarawak) was studied by Lawan et al. (2015) \cite{Lawan2015} using wind data obtained from MMD for four stations. The proposed topological ANN wind speed prediction method was utilized to predict the wind data for the uncovered areas in Sarawak to generate a more precise spatial map. Two ready-made software (i.e. GEPlot and Google Earth) were utilized to build a DEM model. The target site sampled locations including the longitudes and latitudes were generated using the world geodetic system (WGS 84). These sampled locations were then used to extract the DEM using a terrain zonum solution (TZS) tool. The last element, the surface roughness class was developed using GEPlot software. The study for the first time proposed isovent wind atalas mal of Sarawak at heights from 10 to 40 m. The study reveals that the northeast and southwest and coastal regions of Sarawak have better wind potentials. 

\begin{figure*}[!h]
	\centering
	\includegraphics[width=0.8\linewidth,keepaspectratio]{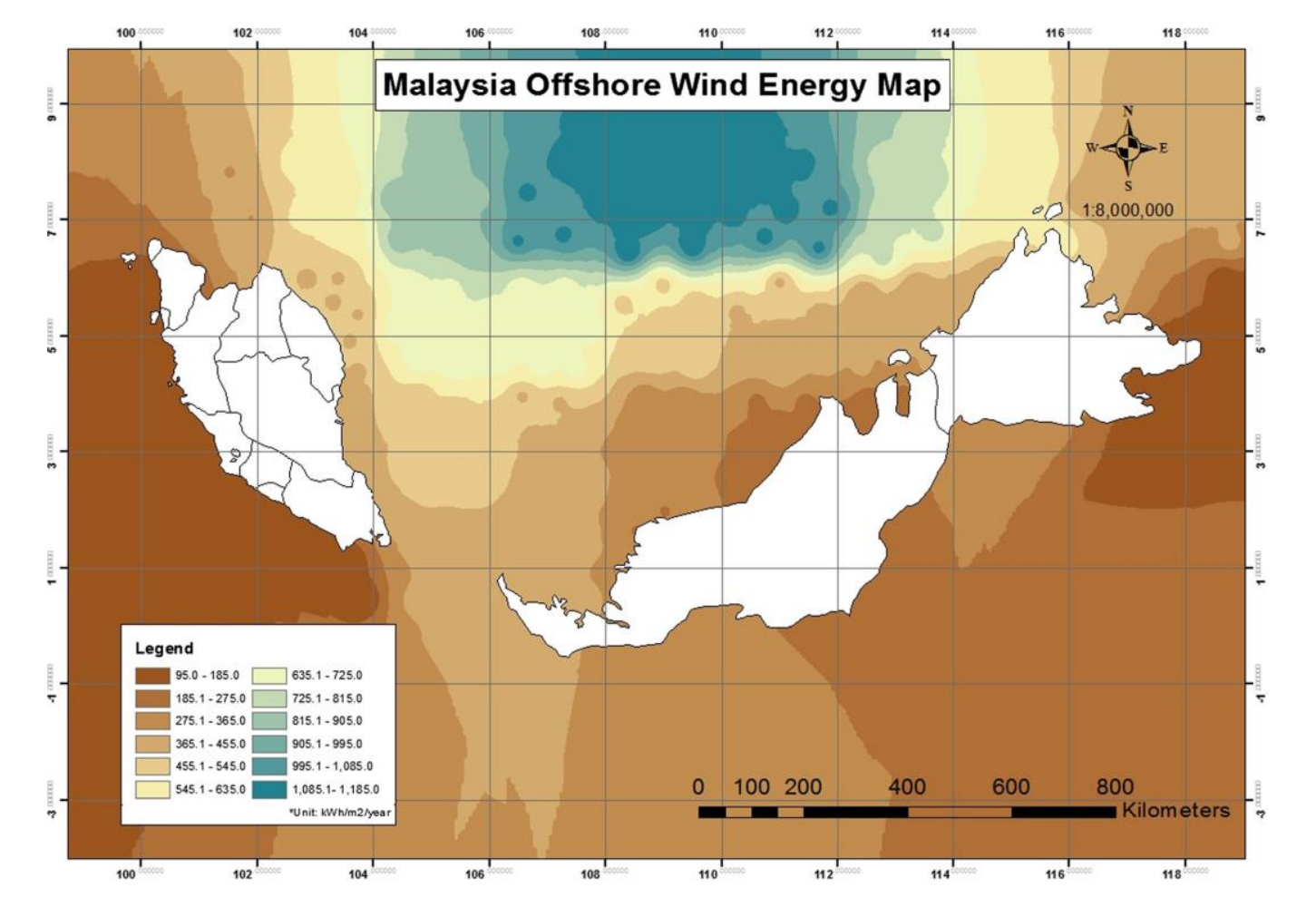}
\vspace{-0.1in}
\caption{Offshore wind energy map of Malaysia. Extracted from \cite{AhmadZaman2019}.}
\label{Fig:Fig4}
\vspace{-0.1in}
\end{figure*}

The main focus of previously reviewed articles was wind resources mapping to explore the wind potential in onshore sites.  Ahmad Zaman et al., (2019) \cite{AhmadZaman2019} studied the mapping and sectorization of offshore wind energy potential locations in Malaysia using multi-mission satellite altimetry data. The satellite altimetry (Jason-1, Jason-2, and Envisat) data was obtained from Radar Altimeter Database Systems located at GNSS and Geodynamics Laboratory, University Technology Malaysia (UTM) for a period of 19 years, from 1993 to 2011. In addition, buoy measurements from two offshore sites were used for validation. Comparing the satellite with buoy data, the average error was found to be 14\% with a correlation of 0.835 and RMSE of 1.99. ArcMap 10.2.2 software using Raster Interpolation with the IDW function was used to generate the annual wind energy maps. Figure~\ref{Fig:Fig4} shows the produces offshore wind energy density map for Malaysia. The study concluded that areas C4 (Terengganu coastline), C9 (Labuan), and D11 (Sabah) have high annual wind speed and wind energy density.

\section{Future research recommendations }
\vspace{-0.1in}
\begin{enumerate}
\item or onshore, several studies have recommended a few potential sites for wind energy harvesting in Malaysia. However, the majority of these studies use the MMD database along with interpolation and extrapolation methods to measure the wind energy feasibility. In order to assess the actual wind speed in these sites, it is important to collect wind data from these sites by installing masts with different heights.

\item Although few studies have conducted offshore wind energy-related analysis mainly using satellite data, the lack of actual offshore wind data makes it difficult to prove the recommendations of these studies. Hence, we also recommended installing wind data collection instruments at the previously highlighted sites in the literature.

\item For optimal sizing of the hybrid energy systems, most of the studies used HOMER software tool, and a rising trend is noticed against the use of modern optimal sizing methods as they offer realistic and promising size optimization. Thus, it is recommended to use such modern methods. For example, Teaching-learning-based optimization (TLBO), the hybrid FPA/SA algorithm, and Non dominated sorting particle swarm optimization (NSPSO).

\item The height of the turbines is noticed to be significantly influencing the wind speed and optimization results. Based on this factor, the constraints in turbines installment, designs, and optimization problems should be considered, instead of using wind speed extrapolation to higher heights which may not reflect the correct wind speeds for different terrains and provide wrong estimates of turbines power output. 

\item The majority of studies concentrate on reliability and cost in the hybrid system, as discussed previously. Few studies focus on the environmental factors such as CO2 emissions. Environmental factors should be considered especially if the energy system consists of a conventional source. 

\item According to the analysed MMD wind data in the previous works, the low wind profile in Malaysia requires a customized wind turbine designs that can operate at low cut-in wind speeds. Therefore, researchers can concentrate more on designing rotor blades that will reach higher rated power at lower wind speed which on the other hand will increase the AEG.

\item The shortage of wind turbines in producing the expected power was probably because of the inappropriate technology used in choosing the wind turbine. Therefore, it is recommended to conduct in-depth research about the wind turbine generator selection process. 

\item There is a need to examine the effect of a different national advisory committee on Airfoils (NACA) profile on the performance of Vatical axis wind turbines (VAWT) since current studies have only considered a single NACA 0012 profile.
\end{enumerate}

\vspace{-0.2in}
\section{Conclusion}
\vspace{-0.1in}
In this review paper, we discussed all wind-related literature studies in Malaysia. Several factors have been ascertained during the analysis, including wind potentiality and assessments, wind speed and direction modelling, wind prediction and spatial mapping, and optimal sizing of wind farms. Despite that the majority of literature studies have paid more attention to the wind potentiality in Malaysia, results analysed are mostly based on theoretical studies shows conflicting conclusions and sometimes grossly inaccurate. However, few studies have shown valuable efforts on studying wind energy even though when using metrological data with limited capabilities. 

According to the wind assessment studies and due to the equatorial location of Malaysia, the wind profile is relatively low in general. However, three potential sites in Malaysia show an acceptable wind profile and could be further investigated for the actual development of wind farms. On average, the reported wind speed is within 2-8 m/s and varies according to the monsoon and site of measurements. Lacking standardization and representation of wind data has led the researchers to use hybrid power systems including other renewable energy sources (i.e. solar and hydropower) along with wind which was highlighted in this review as well.

Several studies have been conducted on wind power energy conversion using various methods viz., power law, computer software (e.g. HOMER), wind turbine power curve, or by using wind power or speed density distributions. Different statistical distributions have been reported and evaluated for wind data in Malaysia with most of the studies suggests the Weibull distribution for wind speed. Mersing (2.4309$^{\circ}$ N, 103.8361$^{\circ}$ E), in many studies, showed the highest wind speed profile with an average above 2.5 m/s which suggests involving small-scale wind turbines with low cut-in speed. Nevertheless, wind power potential for obtained from wind data collected over 10 years, Mersing achieved a power density above 50 W/m$^2$ \cite{Akorede2013, Sopian1995}. Kudat (6.8831$^{\circ}$ N, 116.8466$^{\circ}$ E), however, in other studies (e.g  \cite{Albani2017a}) showed a slightly better wind profile compared to that of Mersing, with average above 2.8 m/s. Persistent analysis of wind speed by  Masseran, et al. (2012)  form Chuping (6.4985$^{\circ}$ N, 100.2580$^{\circ}$ E) showed the lowest variations and the most persistent regardless of the wind speed values. It is worth noting that, some studies consider the annual average of wind speed which doesn't correctly reflect the actual power output as it is heavily dependent on the wind speed fluctuations.

Wind prediction was also a subject of interest in general with a limited number of studies proposed for wind prediction in Malaysia. The conducted review of these studies showed that the proposed methods achieved different degrees of uncertainty lacking to generalized prediction approach which can tackle seasonal and annual wind variations for both short and long-term horizons. On the other hand, the extrapolation of wind speed to different heights could impose significant errors producing falsely estimations of power profiles. The topographic parameters should be considered for both vertical extrapolation and spatial interpolation to produce more accurate wind potentiality assessment analysis. Spatial prediction and mapping of entire or part of Malaysia have been studied in the literature, offering valuable preliminary maps highlighting several potential sites. This is much beneficial for future investigations where before any actual wind turbine installation decisions, the wind data need to be measured to confirm the feasibility of the identified site.

Though most of the previous studies concluded wind power generation for commercial purposes is not feasible in Malaysia, different methodologies were proposed and suggested further analysis of turbine designs, optimal sizing, and hybrid power systems to overcome the significant barriers of wind energy development for low potential wind areas. Most of the wind turbines in the market have cut-in speed above 3 m/s which is beyond the average annual wind speed in Malaysia. The vertical axis wind turbines were highlighted and analyzed in several studies since it can operate regardless of any sudden changes in wind directions. With the rapid growth of technological advancements in wind turbine designs, a fully customized wind turbine designs are still required for such wind profile of Malaysia. In this context, a hybridization of two or more renewable power sources seemed to provide a partial solution to this issue, many studies have investigated this option and concluded the effectiveness of these hybrid systems.
\vspace{-0.2in}

\bibliographystyle{IEEEbib}
\bibliography{references}

\end{document}